# Title

# RELICT: A Replica Detection Framework for Medical Image Generation

## Authors

Orhun Utku Aydin*[1], Alexander Koch*[1], Adam Hilbert*[1], Jana Rieger[1], Felix Lohrke[1], Fujimaro Ishida[2], Satoru Tanioka[1,3], Dietmar Frey[1,4]

* Authors contributed equally to this work

## Affiliations

1 CLAIM - Charité Lab for AI in Medicine, Charité – Universitätsmedizin Berlin, corporate member of Freie Universität Berlin and Humboldt-Universität zu Berlin, Charitéplatz 1, 101117, Berlin, Germany

2 Department of Neurosurgery, Mie Chuo Medical Center, 2158-5 Myojin-cho, 514-1101, Hisai, Tsu, Japan

3 Department of Neurosurgery, Mie University Graduate School of Medicine, 2-174 Edobashi, 514-8507, Tsu, Japan

4 Department of Neurosurgery, Charité – Universitätsmedizin Berlin, corporate member of Freie Universität Berlin and Humboldt-Universität zu Berlin, Charitéplatz 1, 101117, Berlin, Germany

## Corresponding Author Details
Dietmar Frey
dietmar.frey@charite.de

# Abstract

**Background**

Despite the potential of synthetic medical data for augmenting and improving the generalizability of deep learning models, memorization in generative models can lead to unintended leakage of sensitive patient information and limit model utility. Thus, the use of memorizing generative models in the medical domain can jeopardize patient privacy.

**Methods**

We propose a framework for identifying replicas, i.e. nearly identical copies of the training data, in synthetic medical image datasets. Our REpLIca deteCTion (RELICT) framework for medical image generative models evaluates image similarity using three complementary approaches: (1) voxel-level analysis, (2) feature-level analysis by a pretrained medical foundation model, and (3) segmentation-level analysis. Two clinically relevant 3D generative modelling use cases were investigated: non-contrast head CT with intracerebral hemorrhage (N=774) and time-of-flight MR angiography of the Circle of Willis (N=1,782). Expert visual scoring was used as the reference standard to assess the presence of nearly identical replicas. We report the balanced accuracy at the optimal threshold to assess replica classification performance of the analysed methods.

**Results**

The reference visual rating identified 45 of 50 and 5 of 50 generated images as replicas for the NCCT and TOF-MRA use cases, respectively. Image-level and feature-level metrics perfectly classified replicas with a balanced accuracy of 1 when an optimal threshold was selected for the NCCT use case. A perfect classification of replicas for the TOF-MRA case was not possible at any threshold, with the segmentation-level analysis achieving the highest balanced accuracy of 0.79.

**Discussion**

Replica detection is a crucial but neglected validation step for the development of deep generative models in medical imaging. The proposed RELICT framework provides a standardized, easy-to-use tool for replica detection and aims to facilitate responsible and ethical medical image synthesis.

# 1. Introduction

Artificial intelligence for medical imaging has the potential to transform diagnostic workflows and support critical treatment decisions in fields such as radiology *(Saha et al. 2024)*, histopathology *(Dippel et al. 2024)* and dermatology *(Salinas et al. 2024)*. One of the key requirements for developing robust and generalizable deep learning models is access to large, diverse and high-quality training and validation datasets *(Schwabe et al. 2024)*. However, the curation of such medical imaging datasets is often constrained by the challenges of sharing of sensitive medical data *(Legido-Quigley et al. 2025)*, costs of data acquisition and limited availability of expert labeling.

To address these challenges, synthetic medical images of various modalities, anatomical regions and pathological conditions have been successfully generated using generative models such as generative adversarial networks (GAN) *(Ferreira et al. 2024)* or diffusion-based models (DM) *(Kazerouni et al. 2023; Ibrahim et al. 2024)*. Synthetic data can then be used to augment the training set for improved fairness, generalizability, and performance of downstream deep learning models *(Frid-Adar et al. 2018; Khader et al. 2023; Ktena et al. 2024; Khosravi et al. 2024)*. Synthetic data should retain the predictive properties of real data and provide high quality and resolution required for many clinical applications. Thus, synthetic images are expected to be indistinguishable from real images to ensure optimal generative quality and utility *(Park et al. 2021)*.

However, a key challenge in generative modeling is the potential reproduction of real training data at varying levels of similarity. While replication, i.e., synthesis of an identical copy of a training sample, is the major concern, synthetic images can also closely resemble training data without being exact copies *(Akbar et al. 2023; Carlini et al. 2023)*. Higher levels of similarity between synthetic and real images can reduce the privacy benefits of synthetic data and decrease the added value of synthetic data augmentations by limiting diversity. This raises ethical concerns, since data that is not initially publicly shared due to data protection regulations can be unintentionally exposed by releasing a model or publishing synthetic datasets.

The use of generative models in the medical domain is a high-risk application that can jeopardize patient privacy *(Giuffrè and Shung 2023)*. Patient imaging data can serve as uniquely identifiable biometric information, similar to fingerprints *(Packhäuser et al. 2022)*. This vulnerability can be exploited in adversarial attacks, such as membership inference attacks, where information about the inclusion of an individual in a training dataset can be extracted *(Kuppa et al. 2021; Paul et al. 2021)*. Despite the existence of data protection regulations, ethical guidelines and AI research checklists *(Chen et al. 2024; Tejani et al. 2024)*, empirical analysis of memorization in synthetic medical image generation remains largely unexplored *(Ibrahim et al. 2024)*.

Prior works addressed content-based image retrieval *(Gupta et al. 2023)* and memorization in medical image generation using various image similarity measures *(Dar et al. 2024)*. A significant gap in the field is the absence of a standardized tool and methodology for replica detection. Recent works either do not check for memorized replicas at all *(Pinaya et*

*al. 2022; Pan et al. 2023; Peng et al. 2023)* or rely on task-specific, custom approaches for replica detection tailored to individual image generation tasks. For instance, Fernandez et al. calculated the overlap of real and generated labels using the Dice coefficient to find nearest neighbours *(Fernandez et al. 2024)*, Dar et al trained a self-supervised model to project images onto a lower dimensional embedding space and performed replica detection through correlation values *(Dar et al. 2024)*, Packhäuser et al. train a siamese neural network to detect memorized images *(Packhäuser et al. 2022, 2023)*, Akbar et al used correlation of pixel intensities *(Akbar et al. 2023)*, Aydin et al. used a predefined threshold of l2 distance ratio *(Aydin et al. 2024)*. This non-uniformity in the literature warrants a standardized, easy-to-use solution to be used as a validation step in medical image generation research.

In this study, we propose a framework for identifying replicas, near-identical copies of the training data, in synthetic medical image datasets. Our framework evaluates image similarity using three complementary approaches: (1) image level comparison, (2) feature extraction by a medical foundation model, and (3) segmentation-level comparison. We demonstrate our framework on two clinically relevant use cases: generative modelling of non-contrast head CT scans with intracerebral hemorrhage and Circle of Willis arterial segments. By proposing a standard, easy-to-use replica detection framework we aim to contribute to the safe, responsible and ethical deployment of generative models in medical imaging.

# 2. Methods

The data collection for this retrospective study was approved by the local Ethics Committees of following hospitals: Mie Chuo Medical Center institutional review board [permit number: MCERB-202321], Matsusaka Chuo General Hospital institutional review board [permit number: 325], Suzuka Kaisei Hospital institutional review board [permit number: 2020-05], and Mie University Hospital institutional review board [permit number: T2023-7]. Written informed consent was waived due to the retrospective nature of the analysis.

**2.1. Data**
**Use case 1: 3D NCCT with Intracerebral hemorrhage**
In use case 1, the image generation task was to synthesise 3D non contrast computed tomography (NCCT) data with intracerebral hemorrhage (ICH) as the leading pathology. The training dataset included 387 patients with baseline and follow up NCCT imaging within 24 hours (in total 774 images) with the primary diagnosis of ICH from 4 hospitals in Japan: Mie Chuo Medical Center, Matsusaka Chuo General Hospital, Suzuka Kaisei Hospital and Mie University Hospital. The NCCT volumes were registered to the MNI space and all volumes had a shape of 182x218x182 voxels with a voxel spacing of 1x1x1 millimeters. Detailed patient characteristics and dataset information have been previously reported *(Tanioka et al. 2024)*.

A latent diffusion model architecture was used for generative modelling. Latent codes of size (8, 20, 24, 20) were produced using a vector-quantized autoencoder with residual-vector quantization, resulting in 8x downsampling of the original data dimensions. The elucidated diffusion training method was used for training and synthetic images were generated using a DPM-Solver++ for 100 sampling steps *(Karras et al. 2022)*.

An nnUnet segmentation model was trained on the training set of 774 scans with manually segmented binary ICH labels *(Isensee et al. 2021)*. The model was trained for 100 epochs using the default nnUnet hyperparameters, with CT normalization using 5 fold cross validation.

The open source implementation of the diffusion model used in this use case can be found in the following github repository:

*https://github.com/claim-berlin/relict/brain-ae*

**Use case 2: 3D TOF-MRA**

In use case 2, the image generation task was to synthesise healthy 3D time-of-flight magnetic resonance angiography (TOF-MRA) data. A 3D adaptation of the StyleGANv2 architecture was used for generative modelling of the Circle of Willis. The training data was open source and consisted of 1782 3D TOF MRA volumes from 7 different datasets as detailed in a prior work *(Aydin et al. 2024)*. Preprocessing steps included registration to a custom TOF-MRA template, cropping to a region of interested of size 128x128x32 and voxel spacing of 0.62x0.62x0.62 millimeters centered around the Circle of Willis.

An nnUnet segmentation model was trained on 50 patients from the TopCoW dataset to segment Circle of Willis artery segments in a multiclass setting *(Isensee et al. 2021; Yang et al. 2024)*. Paired artery segment labels were merged to create a single class. This resulted in following artery segments: Internal carotid artery (ICA), basilar artery (BA), posterior communicating artery (Pcom), anterior communicating artery (Acom), the posterior cerebral artery (PCA), anterior cerebral artery (ACA), and the first segment of the middle cerebral artery (M1). The model was trained for 1000 epochs using the default nnUnet hyperparameters, with MR normalization using 5 fold cross validation.

The details regarding the generative model architecture, hyperparameters and open source implementation can be found in the following github repository:

*https://github.com/claim-berlin/3D_StyleGAN_Circle_of_Willis*

## 2.2. Replica detection and study design

The definition of replica in scientific literature is ambiguous, can be task-dependent and subjective *(Fernandez et al. 2023; Dar et al. 2025)*. In this work, we refer to a synthetic image as a replica if it is a near identical copy of a real image in the training set and has no distinguishing anatomical or pathological image features compared to the real image.

Our analysis using the proposed replica detection framework consists of the following steps: 1) a generative model is trained on 3D medical imaging data, 2) the trained generative model is used to generate a synthetic dataset, 3) for each synthetic image, training images are ranked based on similarity to identify the closest training image, 4) an expert visual scoring is performed to score similarity of each synthetic image and closest training image leading to a binary replica decision, 5) the various measures are tested at their optimal thresholds for their replica detection performance.

For subsequent automatization of replica detection we propose following steps: steps 1-3 remain the same, 4) ranking of synthetic images within the synthetic dataset by their distance ratios for each measure, 5) visual scoring of a small subset of highest likely replica images with lowest distance ratios 6) visual scoring results are used to optimize use case specific thresholds for automated replica detection in future datasets. The methodological overview of the RELICT framework is shown in Figure 1.

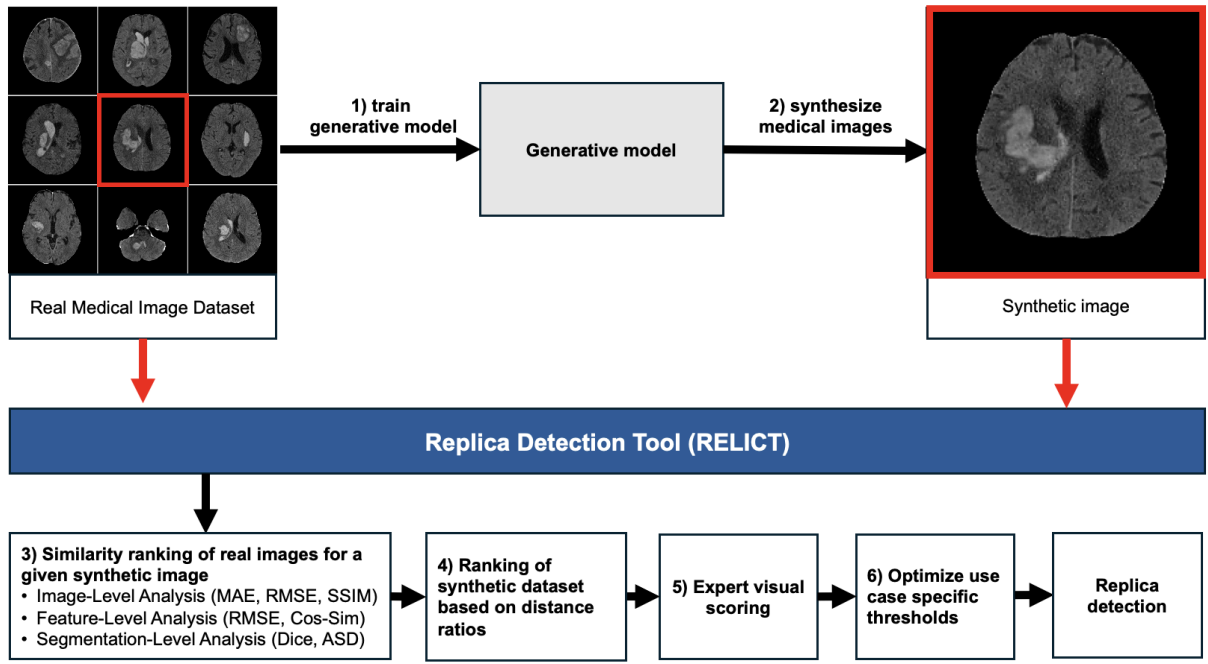

Figure 1. Methodology of the replica detection framework.

## 2.3 Replica detection methodology

A measure of image similarity is required for robust identification of potential replicas in medical imaging. We assess the similarity between a synthetic and real training image from three different analysis perspectives: 1) image-level analysis, 2) feature-level analysis, 3) segmentation-level analysis. The individual measures used in the different analysis levels are described in Appendix 1.

## 2.3.1 Image-level Analysis

In the image-level analysis voxels constituting the synthetic and real image are directly compared based on various measures to compute a distance or similarity score. Image-level analysis constitutes a simple and explainable method to compare images and is adopted in previous works for replica detection tasks *(Carlini et al. 2023; Yoon et al. 2023)*. In our work, we use the following measures for image-level analysis: mean absolute error (MAE), root mean square error (RMSE) and structural similarity index measure (SSIM) *(Wang et al. 2004)*.

### 2.3.2 Feature-level Analysis

Feature-level analysis aims to reduce the dimensions of the original images by using a pretrained encoder, and subsequently comparing feature representations instead of the original voxel or pixel values. The feature-level comparison has been shown to allow a more medically relevant evaluation of image similarity *(Gupta et al. 2023; Jush et al. 2024; Dar et al. 2025)*. In our work we use the pretrained Resnet-50 MedicalNet as a medical foundation model for feature extraction *(Chen et al. 2019)*. The network has been trained on 23 medical image segmentation datasets and has been used in other works to extract feature representations from medical images *(Chen et al. 2019; Tak et al. 2024)*. Feature-level analysis can be computationally faster especially if GPU resources are used for the feature extraction step, since only extracted features are compared instead of whole images. In our work, images are normalized using z-score normalization and encoded using the Resnet-50 pretrained MedicalNet model *(He et al. 2015)*. The dimensions of the resulting feature map is further reduced using an adaptive average pooling layer to a final size of (2048x4x4x4). This embedding is flattened to a feature vector and compared using the RMSE and the cosine similarity.

### 2.3.3 Segmentation-level Analysis

Different medical imaging modalities contain target structures relevant for diagnosis and treatment decisions. For instance, the first use case of this study includes NCCT images of patients with intracerebral hemorrhage. The hemorrhage lesions constitute the region of interest and are more salient for the diagnosis and treatment decisions compared to other parts of the image. Generative models trained on data containing a significant region of interest (ROI) such as hemorrhage lesions should preserve and capture the predictive properties of the real images with respect to the pathology. Therefore, segmentations of ROIs should play a role in replica detection frameworks. Similar ROIs can argue in favour of a generated image being a replica, although there might be considerable differences in the background. In our work, we use the Dice coefficient *(Zou et al. 2004)* and the average surface distance (ASD) *(Yeghiazaryan and Voiculescu 2018)* from an open-source implementation to compare segmentations (https://github.com/google-deepmind/surface-distance).

## 2.4 Distance ratio and replica decision

In our replica detection framework, we use the methodology introduced by Carlini et al. and Yoon et al. *(Carlini et al. 2023; Yoon et al. 2023)*. This approach first computes a measure, such as RMSE, between the generated image under evaluation and all images in the training set. Second, the training images are sorted from most similar to least similar and the closest training image is identified. Third, we use following equation to compute the distance ratio by comparing a synthetic image to real images:

$$M(\hat{x}, x; S_{\hat{x}}) = \frac{M(\hat{x}, x)}{\mathbb{E}_{y \in S_{\hat{x}}}[M(\hat{x}, y)]}$$

for a given distance measure M, where $\hat{x}$ is the synthetic image under evaluation for replica detection, x is the closest image in the training set, $S_{\hat{x}}$ is the subset of n closest images in the training set to $\hat{x}$. This equation computes the measure value for the closest training image x and divides it by the mean value of the n closest training images. This equation is modified from the work by Carlini et al. All experiments were performed with n=50 closest training images for $S_{\hat{x}}$.

The distance ratio provides information about how "abnormally close" the synthetic image is to the closest training image *(Carlini et al. 2023)*. A binary decision whether the synthetic image is a replica can be made by thresholding the distance ratios. Here, the threshold T might depend on image properties (e.g. intensity range, image variation within training set) and should ultimately reflect the user's tolerance for resemblance:

$$\text{Replica Decision} = \begin{cases} \text{Replica, if } M(\hat{x}, x; S_{\hat{x}}) < T, \\ \text{Not a Replica, if } M(\hat{x}, x; S_{\hat{x}}) \geq T. \end{cases}$$

Equation 1 based on the distance ratio was used for the image-level and feature-level analysis. To ensure consistency when applying threshold comparisons, similarity measures were converted into distance-based measures by first normalizing the values between 0 and 1 and subtracting their values from 1 (e.g., a Dice coefficient of 0.7 was transformed into a distance of 0.3).

For the segmentation-level analysis only the absolute value of the segmentation evaluation result was used for replica detection instead of the ratio in Equation 1. This decision was based on the consideration that the segmentation step already isolates the foreground region of interest and disregards background information successfully.

## 2.5. Visual Scoring of Replicas

The respective generative models were used to generate 50 synthetic images for each use case. Since comparing all 50 synthetic images to each training image visually would be infeasible, synthetic images were paired up with a single training image as a preliminary step. For each synthetic image the most similar image from the training dataset was identified based on the RMSE calculated between the synthetic image and each image in the training set. This allows for a more detailed and reliable image comparison within feasible efforts of clinical raters.

Each pair of synthetic and real image was inspected by two senior raters independently and classified based on predefined subjective visual scoring criteria (Table 1). The visual scoring was performed using ITK-SNAP by inspecting the pair of synthetic and real image side by side *(Yushkevich et al. 2006)*. A 4 point Likert-type scale *(Likert 1932)* was chosen to avoid neutral decisions. Visual scores of 3 or 4 led to the classification of a synthetic image as a replica. In cases where the raters disagreed on the replica classification decision, the images were re-evaluated by each rater individually using the same criteria. The visual scoring is illustrated with examples in Figure 2.

| Scale | 1 | 2 | 3 | 4 |
|---|---|---|---|---|
| **Point** | **Certainly not a Replica** | **Probably not a replica** | **Probably a Replica** | **Certainly Replica** |
| **Description** | Two different images, no resemblances | Images are different, with considerable differences in anatomy, pathology or background | Images are very similar, with some minor differences in anatomy, pathology or background | Images are mostly identical |

**Table 1.** Subjective visual scoring for replica detection ground truth creation. The scoring is performed using a Likert-type scale where increasing scores reflect the rater's subjective confidence that an image is a replica.

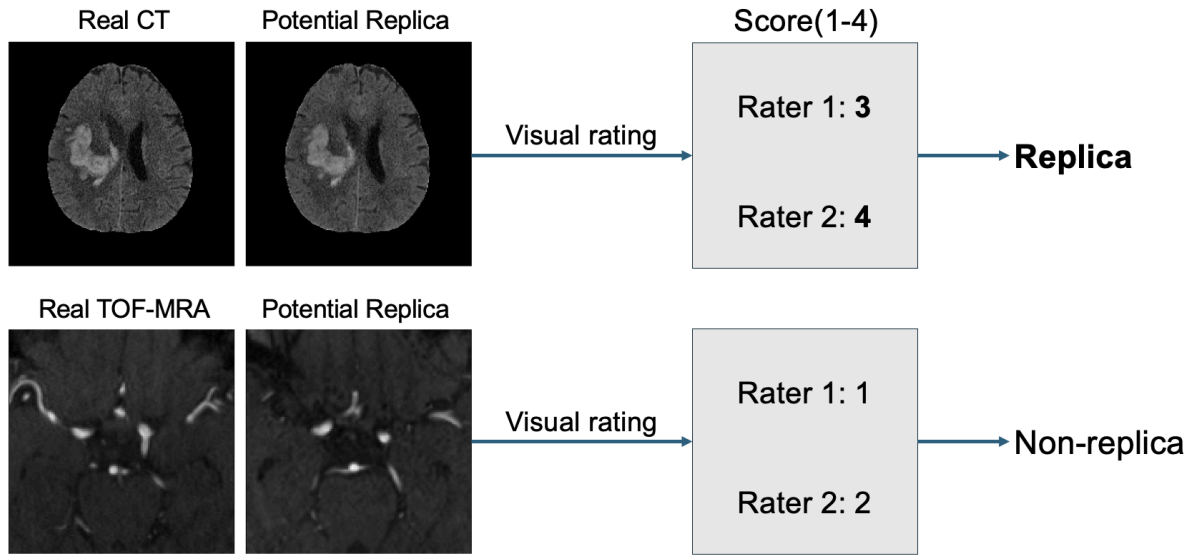

**Figure 2.** Subjective visual scoring examples for two pairs of real and generated images.

### 2.6. Software

The framework is written in the Python programming language (Version: 3.10) and implements the image comparison measures from open-source libraries. The authors will make efforts to assess ongoing research in the field and aim to keep the repository updated with successful replica detection methods as they are developed. Additionally, we welcome contributions from the open-source community to expand the framework.

Computation was performed on the HPC for research cluster of the Berlin Institute of Health using 50 CPU cores and a single V100 GPU. The replica detection code is available open-source in the following github repository:

*https://github.com/claim-berlin/relict*

### 2.7 Evaluation

For each analysis, the replica detection thresholds were analyzed by systematically evaluating performance across 0.01 increments relative to the value of the measure. The performance was reported using balanced accuracy to equally consider sensitivity and specificity. The runtime was reported in minutes for each analysis method, using a typical workstation computer.

# 3. Results

**3.1 Visual rating results**

In the NCCT use case, raters identified 45 out of 50 synthetic images as replicas, with the remaining 5 classified as non-replicas. In the TOF-MRA use-case, 5 images were identified as replicas and 45 as non-replicas. The raters agreed in the replica detection decision in 46 out of 50 cases (92%) for the NCCT use case and 41 out of 50 cases (82%) for the TOF-MRA use case. The median visual score was 3 and 4 for the NCCT images and 1 and 2 for the TOF-MRA images for the two raters respectively. Figures 2-4 present example pairs of real and synthetic images assessed by the raters.

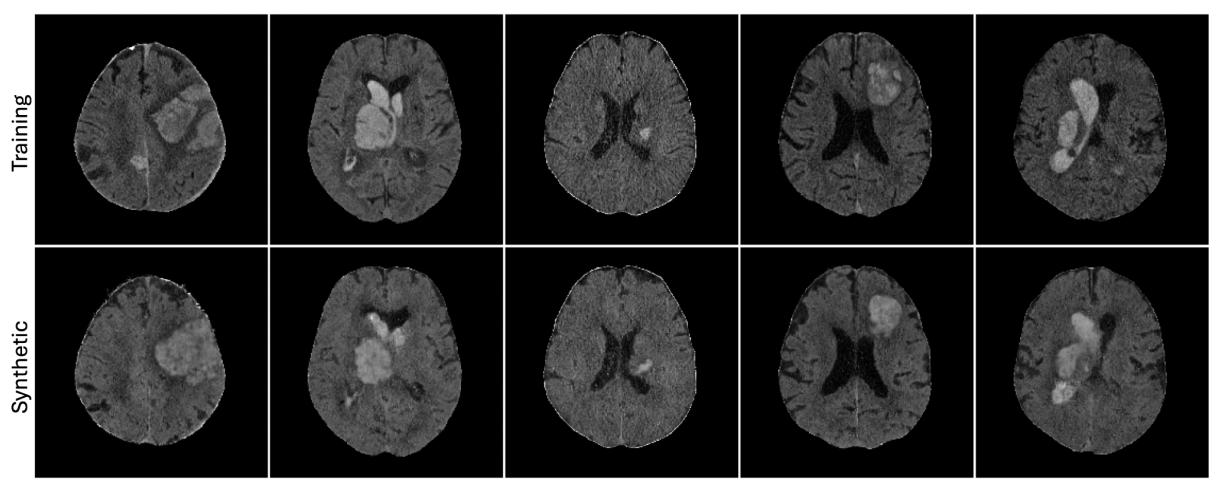

**Figure 2.** NCCT images classified as non-replicas based on visual rating.

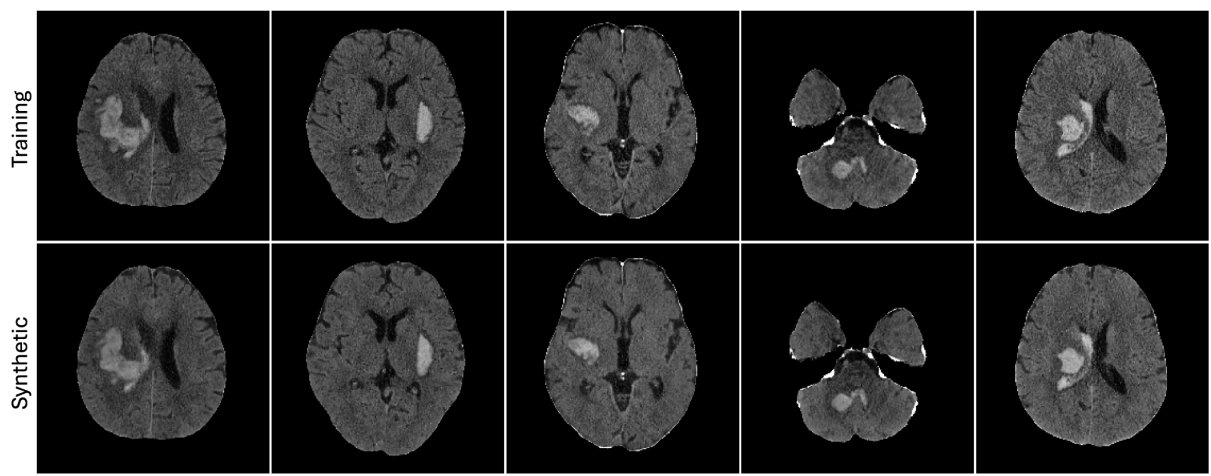

**Figure 3.** NCCT images classified as replicas based on visual rating.

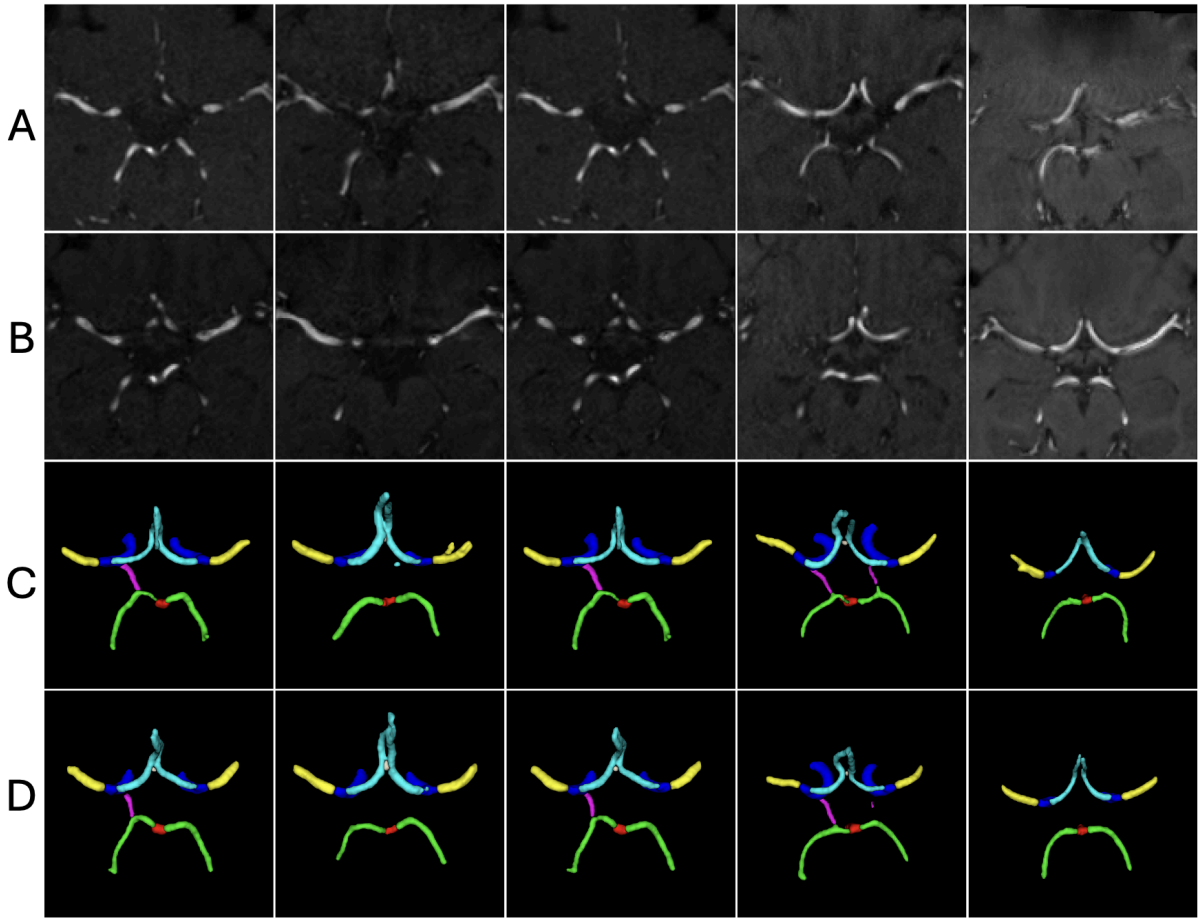

**Figure 4.** TOF-MRA images classified as replicas based on visual rating. A. Real TOF-MRA data, B. synthetic TOF-MRA data, C. segmentations of real volumes from row A, D. segmentations of synthetic volumes from row B.

### 3.2 Replica detection results

The optimal threshold for replica detection varied based on the analysis-level and measure used (Figure 5, Appendix 1B). For the NCCT use case, all measures tested on the image-level and feature-level analysis allowed for finding an optimal threshold to identify replicas. This indicated a perfect alignment with the visual rating (balanced accuracy of 1). The segmentation-level measures Dice and ASD had lower balanced accuracies of 0.96 and 0.98 respectively. The closest images identified by RMSE based preselection were also identified by all other measures as the closest image in 47 out of 50 images.

In the TOF-MRA use case, the analysed measures ranked training images differently for all 50 synthetic images and thus identified different images as closest, compared to the preselection method. An example synthetic image with different closest training images identified by RMSE, feature cosine similarity and multiclass ASD are shown in Figure 6. The segmentation-level ASD measure enabled the highest replica detection performance with a balanced accuracy of 0.8 and 0.79 in binary and multiclass settings. All analysed image-level and feature-level measures resulted in a balanced accuracy lower than 0.72.

An overview of all used measures with their respective analysis levels and runtimes can be found in Table 2.

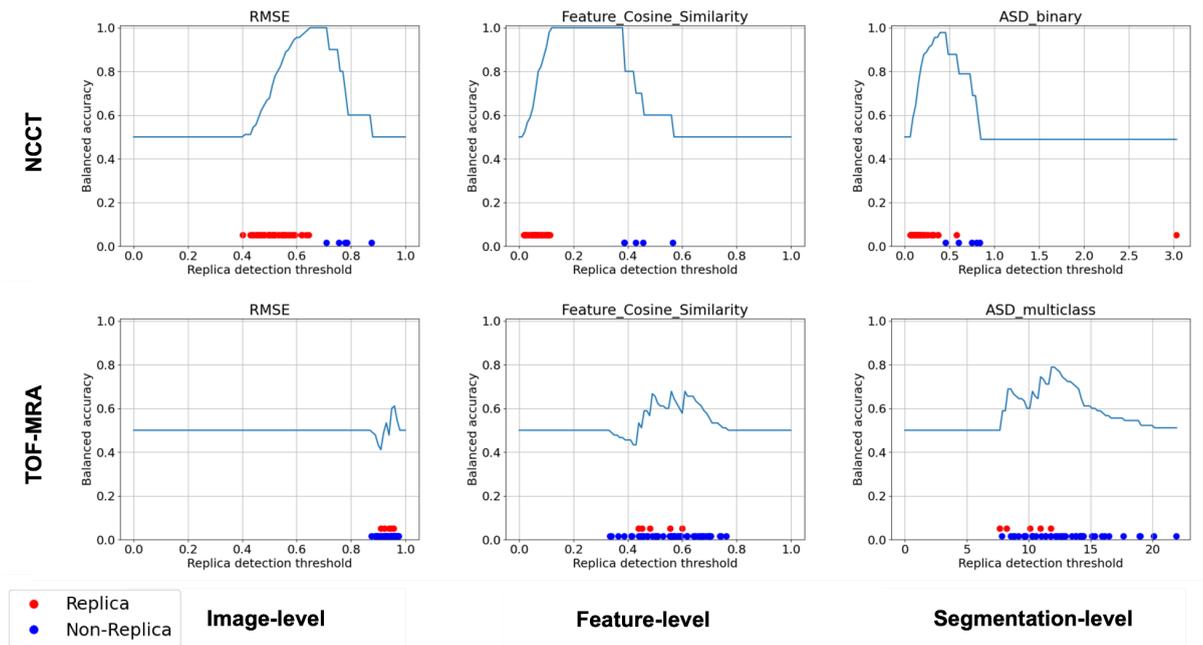

**Figure 5.** Quantitative evaluation of replica detection performance for NCCT and TOF-MRA use cases.

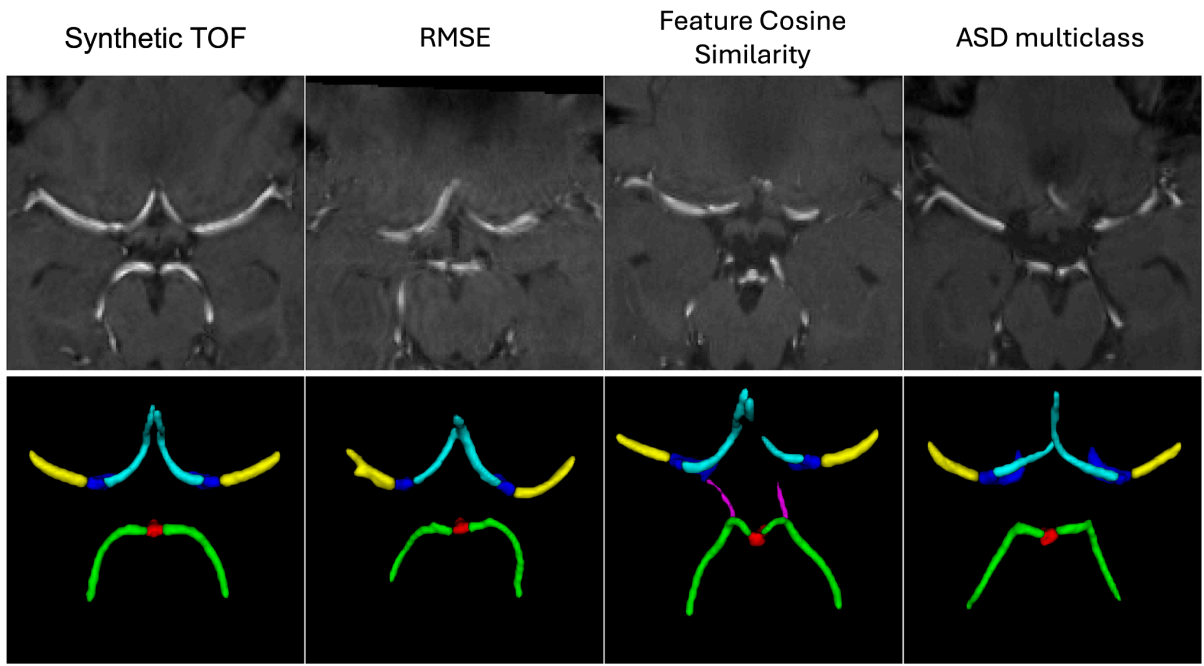

**Figure 6.** Closest training images of generated TOF images from the training dataset selected by various measures.

**Table 2.** Replica detection methods overview and runtime analysis.

| Measure | Analysis Level | Runtime | |
|---|---|---|---|
| | | **NCCT** | **TOF-MRA** |
| **Mean Absolute Error (MAE)** | Image-level | 16 mins | 2 mins |
| **Root Mean Square Error (RMSE)** | Image-level | 16 mins | 2 mins |
| **SSIM** | Image-level | 76 mins | 11 mins |
| **Embeddings RMSE** | Feature-level | 4 mins | 2 mins |
| **Embeddings Cosine similarity** | Feature-level | 4 mins | 2 mins |
| **Dice (binary)** | Segmentation-level | 23 mins | 3 mins |
| **Dice (multiclass)** | Segmentation-level | - | 4 mins |
| **Average surface distance (binary)** | Segmentation-level | 31 mins | 16 mins |
| **Average surface distance (multiclass)** | Segmentation-level | - | 16 mins |

# Discussion

We propose a replica detection framework to be used in medical image generation research as a standard model validation step. Memorization in generative models and resulting replicated images can be detected using image-level, feature-level or segmentation-level comparisons. Our research confirms reports of memorization in medical and natural image generation studies, and aims to raise awareness about the significant risks posed by this underexplored issue.

Beyond confirming the threat of image replication in medical image generation, our analysis yielded more insights into the challenge. Visual rating revealed a clear discrepancy in the percentage of replicated images in the two use cases; 90 percent of all generated NCCTs were identified as replicas in comparison to only 10 percent in the TOF-MRA use case. The image-level and feature-level analysis agreed best with the visual scoring in the NCCT use case, reflecting the near-identical appearance of the volume pairs even when evaluated on individual axial slices. In contrast, for the TOF-MRA use case, the segmentation-level analysis outperformed the image- and feature-level methods. This is likely due to images showing more abstract similarities, i.e in vessel anatomy, bifurcations, that are not fully captured by image or feature level comparisons. Furthermore, we demonstrated utilization of the provided ranking of synthetic images based on their distance ratios, enabling the identification of high likely replica candidates automatically. Decision thresholds can be defined and tuned by subjective visual assessment and desired criteria for similarity to serve automated detection in subsequently generated synthetic datasets.

Memorization is a major concern and has implications for model development, downstream use of synthetic data, and data sharing. Memorization is being increasingly considered during the development of generative models and factors causing it are being explored to develop mitigation strategies *(Somepalli et al. 2023; Dutt et al. 2024; Dombrowski et al. 2024)*. Research in natural image generation shows that models trained on smaller training dataset sizes are more prone to memorization. Additionally, it has been reported that larger, more complex models memorize faster *(Tirumala et al. 2022)*. In the NCCT use case, we trained a large diffusion-based model on a small dataset of 774 images with high resolution in 3D. In addition, the dataset contained baseline and follow-up images of the same patients, with a high degree of similarity of the images. This might make memorization more probable since data duplication has also been reported to increase memorization *(Carlini et al. 2023)*.

Recently, differential privacy (DP) has been proposed as a method for generative image modelling to share sensitive patient data with known guarantees about privacy-preservation. Differential privacy is an active area of research and allows a tradeoff between data utility and preservation of privacy. By decreasing the parameters (epsilon, delta), it is possible to obtain a stronger privacy guarantee, at the cost of reducing the utility of the data. Thus, practitioners always have to decide between potentially leaking sensitive data but providing high utility or vice versa. Especially for diffusion based models the implementation of differential privacy is practically challenging *(Dockhorn et al. 2023)*. We see this as an important example application of our replication detection pipeline as a filter to additionally safeguard sharing data with potential replication of patient information. Since DP can not guarantee that no sensitive data is leaked, it is also not a panacea to avoid legal constraints in data sharing.

Generated data is increasingly being used for downstream data augmentation in medicine *(Ktena et al. 2024)*. Memorized images are highly unlikely to benefit downstream tasks since they do not provide additional diversity to a given training set. We hypothesize that our framework can be used as a filter to select valuable and unique images among many generated images. Data augmentation using more unique images can help address the data diversity problem in medical image deep learning models (Hofmanninger et al. 2020). With this proposed replica filter, the added value of generative models - even predominantly memorizing models as in the NCCT use case - can be tested in downstream tasks.

Based on our findings we have several recommendations for using RELICT for replica detection. First, the replica detection threshold varies by dataset and measure used to compare images. Thus, instead of relying on a single threshold, we recommend using the proposed framework to identify pairs of synthetic and real data that are most likely to be replicas. This information is an output of the framework where pairs are ranked based on either their distance ratios or segmentation measure values. In a separate step, the ranking of most likely memorized images should be manually inspected to optionally fine-tune a replica detection threshold for a given dataset if automation is desired. Second, the training setup of the generative model should be considered during replica detection. Especially if the generative model was trained using data augmentations, RELICT cannot guarantee to find the closest real image, since augmentations are not considered explicitly. In this case, feature-level analyses can be explored such as training of self-supervised models using contrastive learning for replica detection for individual datasets *(Dar et al. 2024)*, although these approaches require training of a dataset specific model. Alternatively, replica detection methods that increase robustness towards variations caused by data augmentation could be explored for medical imaging *(Somepalli et al. 2023)*. Third, if the training dataset is large, image-level analyses can be computationally very expensive and therefore cosine similarity of embeddings using an established medical foundation model, such as MedicalNet should be preferred. Future additions to the framework can be made using more recent foundation models such as BiomedParse *(Zhao et al. 2025)* or MedSAM *(Ma et al. 2024)*.

Several checklists and reporting guidelines aim to ensure reproducibility and transparency of AI research in medicine and to provide reliability of published scientific evidence *(Tejani et al. 2024; Lekadir et al. 2025)*. Current guidelines however, are mostly concerned by predictive AI performance and reporting and are not specifically adapted for potential shortcomings of generative models. Our results suggest that a standardised replica detection framework can be used to reveal replicas in synthetic datasets and thus provides insights to reliability and quality of synthetic datasets. We aim to raise awareness and caution on sharing of synthetic data, and advocate for inclusion of replica detection requirements as part of reporting guidelines on generative AI studies.

Our study has several limitations. First, the proposed replica detection framework could only be tested in a limited number of radiological use cases, each using a single generative modelling approach, hence the memorization could not be directly compared between model architectures. This was due to the fact that successfully training a generative model requires extensive hyperparameter tuning and computational resources. Second, a limited number of image comparison measures were used and only a single medical foundational model was tested due to the exploratory nature of our replica detection tool. Third, in the subjective visual rating the two senior raters assessed only the closest image suggested by a single measure, RMSE, because pairwise comparison of all images in the training set was not feasible due to time constraints.

## Conclusion

Replica detection is an important, but often neglected quality assurance step for validation of generative models in medical imaging. Standardized replica detection methods need to be developed and included in AI in radiology and medicine checklists. Replica detection methods are a crucial element for detecting patient privacy violations of generative models. Our developed framework provides an important step towards standardized and rigorous validation practices of generative models with potential for safer sharing of synthetic medical image data.

# Data availability statement

Data are available from the corresponding author upon reasonable request.

# Acknowledgements

Computation has been performed on the HPC for Research cluster of the Berlin Institute of Health.

# Disclosures

The authors declare that they have no conflicts of interest.

# Funding statement

This work has received funding from the European Commission through Horizon Europe grant VALIDATE (Grant No. 101057263, coordinator: DF) and Horizon Europe Grant CYLCOMED (Grant No.101095542). In addition, This work has received funding from the German Federal Ministry of Education and Research (ANONYMED Project 16KISA043).

# Author contribution statement

Orhun Utku Aydin: Writing – review & editing, Writing – original draft, Visualization, Software, Methodology, Investigation, Formal analysis, Data curation, Conceptualization. Alexander Koch: Writing – review & editing, Visualization, Software, Investigation, Formal analysis, Data curation, Conceptualization. Adam Hilbert: Writing – review & editing, Visualization, Supervision, Software, Project administration, Methodology, Investigation, Funding acquisition, Formal analysis, Conceptualization. Jana Rieger: Writing – review & editing, Visualization, Software. Felix Lohrke: Writing – review & editing, Methodology, Software. Satoru Tanioka: Writing – review & editing, Writing – original draft, Visualization, Validation, Formal analysis, Data curation. Dietmar Frey: Writing – review & editing, Writing – original draft, Validation, Supervision, Formal analysis, Resources, Project administration, Investigation, Funding acquisition.

# Supplementary Materials

## Appendix 1.

### A. Measures for image comparison

**Image-level analysis**

**Mean Absolute Error**

The Mean Absolute Error (MAE) is calculated as the average of the absolute differences between corresponding voxels in the generated and real volumes:

$$MAE = \frac{1}{N} \sum_{i,j,k} \left| x_{i,j,k} - y_{i,j,k} \right|$$

where N is the total number of voxels.

**Root Mean Squared Error**

The Root Mean Squared Error (RMSE) is calculated as the square root of the average of the squared differences between corresponding voxels in the generated and real volumes:

$$RMSE = \sqrt{\frac{1}{N} \sum_{i,j,k} \left( x_{i,j,k} - y_{i,j,k} \right)^2}$$

**Mean Structural Similarity Index Measure**

The SSIM was calculated using the structural_similarity function of the skimage.metrics library based on the implementation by Wang et al using a gaussian weighting function with a standard deviation of 1.5 in the calculation of SSIM *(Wang et al. 2004)*. The SSIM considers the luminance, contrast and structure of the images. To provide a single measure for comparison between a pair of real and generated images the mean SSIM was used for replica detection *(Zhou Wang and Bovik 2009)*.

$$\text{SSIM}(x,y) = \frac{(2\mu_x\mu_y + C_1)(2\sigma_{xy} + C_2)}{(\mu_x^2 + \mu_y^2 + C_1)(\sigma_x^2 + \sigma_y^2 + C_2)}$$

Where μx, μy are the pixel mean values, σx, σy are the standard deviations, σxy the covariance of x and y, and C1, C2 are stabilisers.

**Feature-level analysis**
**Cosine similarity**
The cosine similarity was calculated after flattening the 3D images to 1-dimensional vectors. The numpy library was used for calculation with the following formula.

$$Cosine\ similarity = \frac{\mathbf{x} \cdot \mathbf{y}}{\|\mathbf{x}\|\|\mathbf{y}\|}$$

where x . y is the dot product and ||x|| denotes the euclidean norm.

**Segmentation-level analysis**

**Dice coefficient**

The Dice coefficient is arguably the most popular measure for segmentation performance assessment. We use the Dice coefficient to perform a region of interest level analysis by comparing segmentations of the leading structures in generated and real images. The Dice coefficient can be calculated using the confusion matrix values of true positives (TP), false positives (FP) and false negatives (FN).

$$\text{Dice} = \frac{2TP}{2TP + FP + FN}$$

**Average Surface Distance**
The average surface distance is the average distance of outline of the predicted surface to the outline of the ground truth surface and vice versa.
Let S1, S2 be two surfaces and let d(s, S) be the distance from voxel s to surface the S. The distance is defined as

$$d(s, S) = \min_{s' \in S} \|s - s'\|$$

then the average surface distance ASD can be defined as:

$$ASD(S1, S2) = \frac{1}{|S_1| + |S_2|} \left( \sum_{s_1 \in S1} d(s_1, S_2) + \sum_{s_2 \in S2} d(s_2, S_1) \right)$$

In the exception case for the multiclass evaluation, where one class has an empty segmentation, the 95 th percentile Hausdorff distance of the whole image was used as the ASD value for that class.
For our replica detection framework, we use an open source implementation of ASD from (https://github.com/google-deepmind/surface-distance) due to its popularity and ease of integration. Other distance based measures might be used interchangeably to assess similarity between segmentations in the scope of replica detection.

## B. Full replica detection plots

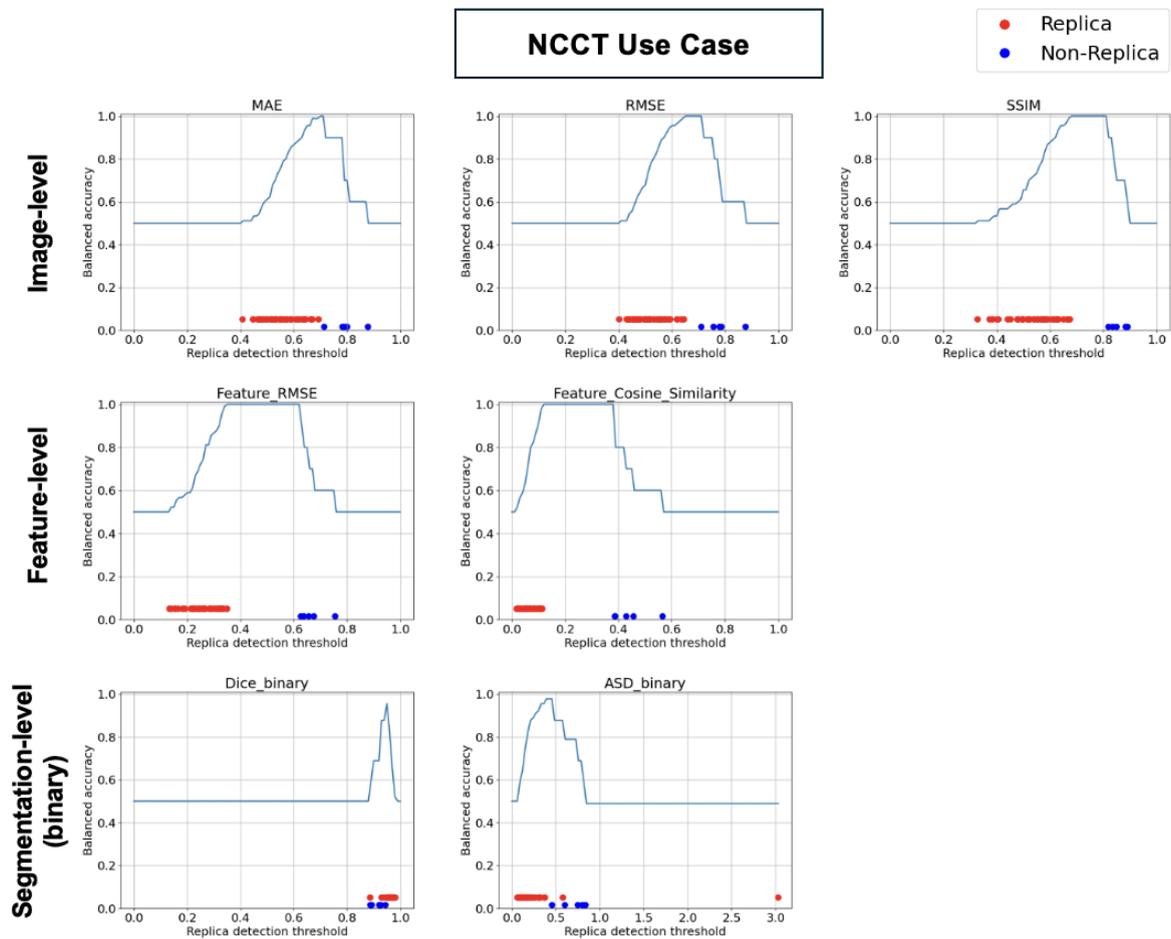

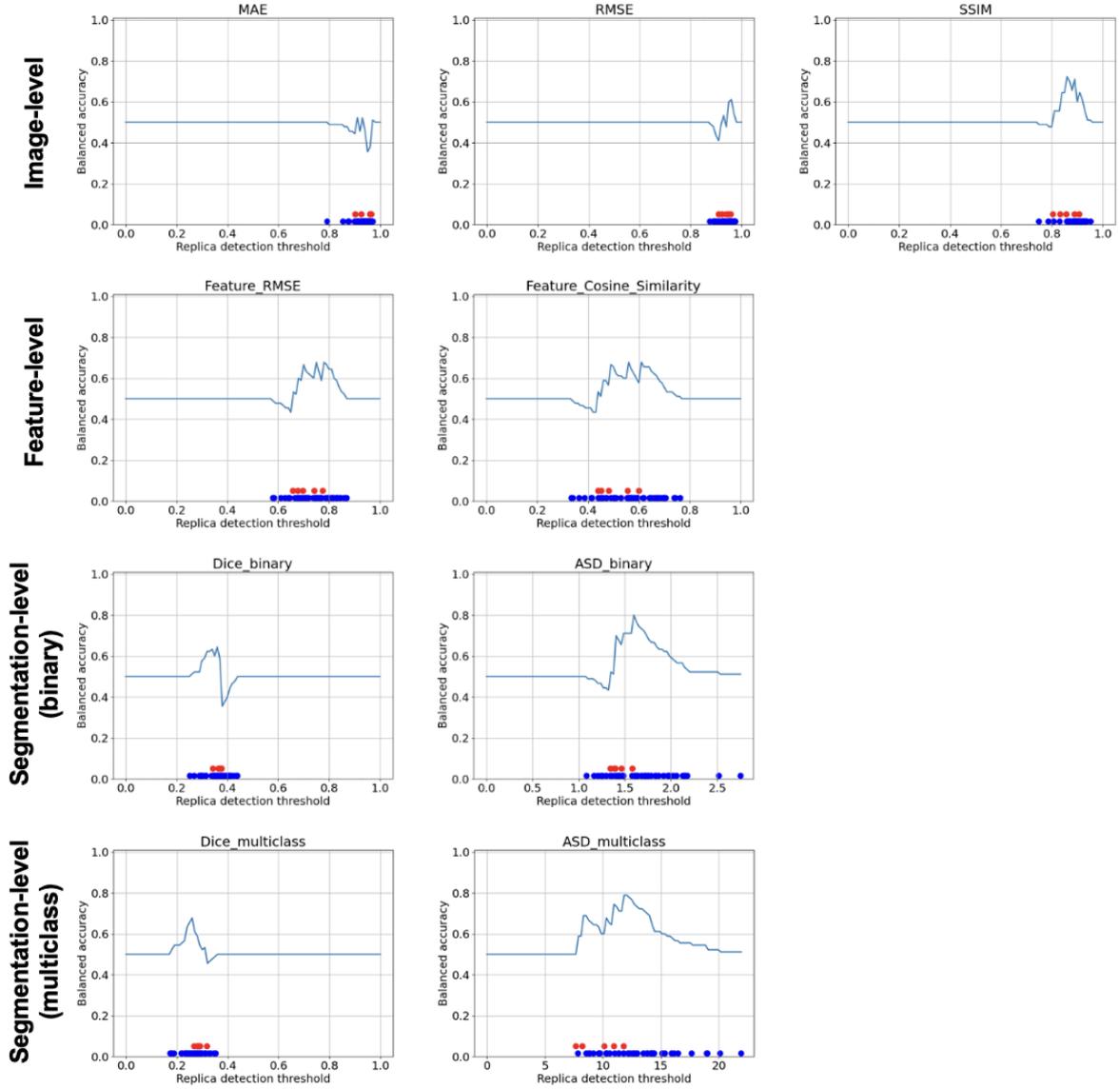